# Genetic recombination as DNA repair

Dmitri Parkhomchuk[1*], Alice C. McHardy[1], Alexey Shadrin[2]

[1]Department for Computational Biology of Infection Research, Helmholtz Center for Infection Research, 38124 Braunschweig, Germany

[2]Alacris Theranostics GmbH Berlin, Germany

*Correspondence to: Dmitri.Parkhomchuk@helmholtz-hzi.de

**Abstract**
Maintenance of sexual reproduction and genetic recombination imposes physiological costs when compared to parthenogenic reproduction, most prominently: for maintaining the corresponding (molecular) machinery, for finding a mating partner, and through the decreased fraction of females in a population, which decreases the reproductive capacity. Based on principles from information theory, we have previously developed a new population genetic model, and applying it in simulations, we have recently hypothesized that all species maintain the maximum genomic complexity that is required by their niche and allowed by their mutation rate and selection intensity. Applying this idea to the complexity overhead of recombination maintenance, its costs must be more than compensated by an additional capacity for complexity in recombining populations. Here, we show a simple mechanism, where recombination helps to maintain larger biases of alleles frequencies in a population, so the advantageous alleles can have increased frequency. This allows recombining populations to maintain higher fitness and phenotypic efficiency in comparison with asexual populations with the same parameters. Random mating alone already significantly increases the ability to maintain genomic and phenotypic complexity. Sexual selection provides additional capacity for this complexity. The model can be considered as a unifying synthesis of previous hypotheses about the roles of recombination in Muller's ratchet, mutation purging and Red Queen dynamics, because the introduction of recombination both increases population frequencies of beneficial alleles and decreases detrimental ones. In addition, we suggest simple explanations for niche-dependent prevalence of transient asexuality and the exceptional asexual lineage of *Bdelloid rotifers*.

**Introduction**
In comparison with parthenogenic (or clonal or asexual) reproduction, recombination and sexual reproduction require the increase of complexity in a number of phenotypic features for mating partner choice, mating itself and recombination, *e.g.* by providing the corresponding (molecular) machinery. This creates additional costs for the organisms and must therefore provide some compensating advantages if it is not to be lost through selection. To explain these advantages, numerous hypotheses have been proposed (Maynard Smith 1978; Kondrashov 1993; Hartfield and Keightley 2012). They largely fall into two big groups (West et al. 1999; Meirmans and Strand 2010): The first one focuses on the effect of (deleterious) mutations at individual loci accumulated over time and is proposed by population geneticists (Mutational Deterministic hypothesis and Fisher-Muller hypothesis). These models were motivated by the mathematical formulations of classical population genetics and have emphasized respective parameters like mutation rate, population size, allele frequencies and interactions between (deleterious) mutations at different loci in the face of selection (epistasis). The second group focuses on the change of environmental conditions, and thus fluctuations in selection, as a cause for the maintenance of sex and recombination. These models were mainly proposed by evolutionary ecologists--initially presented as verbal arguments, they have also been formalized using certain parameters, *e.g.* virulence in host-parasite interactions (Red Queen dynamics (Bell 1982; Lively and Morran 2014)). As neither of these schools was able to provide an explanation consistent with all empirical data, attempts have been made to unify them (West et al. 1999), coupled to an urge for parameter estimation for model validation. However, it has also been argued that most of the parameters in these models are either hard to determine or it is unclear how relevant they actually are (Butlin 2002) and no universal explanation of the maintenance of sex and genetic recombination has been found.

However, genetic recombination is ubiquitous among organisms and its principle is quite simple: inter-individual DNA exchange for progeny generation. Thus, the demonstration of its basic advantage should also be simple and robust, when represented in a proper "coordinate system". Despite the decades of efforts, such demonstration has not been provided. We hypothesize that the reason for this is that the traditional population genetic framework is lacking the formalism for organismal complexity dynamics, and here we attempt to address this omission.

Our framework (Shadrin et al. 2013; Shadrin and Parkhomchuk 2014) is not based on previous population genetics models, but is constructed from scratch to allow the quantification of species genomic information and hence its phenotypic complexity and efficiency. While selection based on nucleotide weights is a common modeling approach, the main novelty of our model is that a population is operating at its "error threshold" making the model nearly parameters-free, for example a population size plays no role in effects we describe. This error threshold puts limits on attainable population fitness and introduces a "motivation" for a population to lessen the errors load by various means, forestalling we hypothesize that recombination is one of them, while, for example, DNA repair is the other.

The role of this threshold in evolution can not be overemphasized. In 1930 (notably before Information Theory) Fisher addressed a similar question of evolution limits using, a "geometrical model" or "microscope analogy". It is apparently difficult to make a perfect image with a microscope with multiple knobs, when each knob influences multiple image parameters (pleiotropy and epistasis) and knobs are adjusted randomly. In computer science that is known as "dimensionality curse". For simplicity, in our model there is no pleiotropy, but instead each knob modifies one feature (QTL) and at each step (generation) few knobs are adjusted randomly and this combination is evaluated by selection. In this case perfect fitness is difficult to attain because when one knob was moved in a "good" direction, others might go in "bad" directions. In this scenario a population average fitness is not perfect, though rare individuals might be close to it. Notably, now it is not inasmuch a "dimensionality curse" but a "noise curse"--the central problem addressed by Information Theory (IT). Here we use this model to formalize species complexity dynamics.

Hence, we can directly approach the question how recombination compensates for the burden of maintaining more complexity by providing the extra capacity for building up and maintaining organismal complexity. Here, we show the general phenomenon that sexually reproducing (or just randomly recombining) species can compensate for the burden and can maintain an extra complexity and thus increased efficiency. How this extra efficiency translates into a phenotype, for example better energy efficiency, better resistance to parasites, faster reaction, sharper teeth, better memory, and so on, is a niche-specific question and is not important for our considerations. In most cases, the increased capacity will have to be larger than the expenditures on recombination maintenance, hence providing the ubiquitous advantage of sexual reproduction in particular for highly evolved species in complex saturated niches. However we also show some plausible scenarios for the persistence of optional parthenogenic reproduction.

Information-theoretical terms, are given to connect the model with IT mathematical concepts and structures. They are not required for understanding of the model workings. The crucial difference of our model from others, are in starting assumptions of high functional mutation rates, which are observed in multicellular organisms, including humans: high mutation rate opposes selection efforts to gain a "perfect" genome. So a balance is established: selection pushes fitness up, while random mutagenesis degrades it. We show that recombination affects this balance, allowing for higher fitness. For bacteria, where mutation rates can be lower than one per-genome per-generations, other sources of stochasticity should be considered, such as frequent fluctuations of environment, which continuously make a genome "imperfect", so the balance should be explicitly introduced there too. This scenario is not covered in this paper. We claim that without this "balance" considerations, it is impossible to reveal the effects of recombination. Other models do not introduce such balance explicitly, the closest analogy is "error threshold" of quasispecies models, but there are important differences from our model (Shadrin and Parkhomchuk 2014).

**Methods**
It proved to be difficult to formalize the information content of a whole single genome, because the "context" of this information is unknown *a priori*. For example, the same nucleotide sequence, put in different conditions (cellular/phenotype/environment), can have different functionality, it can be beneficial, useless or detrimental. However, if we define a genotype-phenotype mapping for an entire species, or, more accurately, a boundary of this mapping, there is an easy way to formalize the information content for this species as a whole with regard to its environment. This provides an estimate of species complexity.

Consider a species with a genome of length *L*. Ignoring, for simplicity, a chromosomal structure, as if a genome is a single chromosome, we can construct the set $N=4^L$ of all possible genomes of this length. However, only a small fraction of these genomes will belong to our given species, we denote the total number of such genomes $N_s$. How much information does this species then contain? A classical way to define information is a number of bits, yes/no responses, a receiver needs to obtain from the sender of information. For example, if only a single genome fits into an overly strict definition of a species, the information is $\log_2(N)=2L$, so each nucleotide position provides 2 bits of information. The receiver could recursively bisect the set *N* of all possible genomes, and ask the sender if a given half contains the species phenotype. After $\log_2(N)$ steps (on average) the receiver would get to the phenotype. However a species definition usually accommodates for some genetic variability in a population, so that much more than one genome would suit a given species definition. In this case the information content of the species is smaller: $\log_2(N/N_s)$, because now the receiver can lock into the species phenotype faster than through bisecting, using correspondingly smaller patches for querying. In an imaginary limiting situation when any genome of *N* is the species, the information is zero because the phenotype is so "unspecific" in this case that any genome suits the species definition and the receiver can guess the species correctly at once. Loosely speaking a "degree of specificity" among all possible genomes is a measure of species information.

Now, we apply this coordinate system to describe a species' genetic composition, keeping the model as simple as possible. Let us assume equal abundance of all nucleotides, no interactions of loci or variants (no epistasis) and the only mutations occurring are single base substitutions. In this case, a large equilibrium population (>>$N_s$), which contains all possible species-specific genomes can be used to stack genomes vertically (similar to a multiple sequence alignment) and to construct the species' genomic "sequence logo" (Schneider and Stephens 1990). Each nucleotide position in this logo is represented by a length 4 vector of nucleotide frequencies in the population. The genetic information (*GI*) content of a position is defined through nucleotide frequencies as

$$GI(P) = 2 + \sum_{N \in \{A, G, C, T\}} f_N \log_2 f_N \quad \text{(Eq. 1)},$$

and the total genomic information is the sum over all positional *GI values*.

The definition of *GI* (Eq. 1) is not arbitrary, it can be interpreted as "positional information content". It reflects the "degree of specificity" of molecular interactions and its mathematical properties ensure the additivity of *GI* values (Shadrin et al. 2013). As we have previously shown, the $N\_s$ can be expressed as $2^{(2L - GI\_total)}$ due to Asymptotic Equipartition Property, because that is the size of the typical set which is generated by a given *GI*-profile (the set of frequency vectors of all genome positions). For a fixed environment and fixed population parameters, the species typical set of possible species genomes ($N\_s$) is constant by definition and so is the total *GI*. Here, we mainly address the effects of recombination in equilibrium populations and demonstrate that in this context, it has ubiquitous advantages under plausible assumptions.

A species with its environment is completely defined by the $N\_s$ set, and the corresponding whole genome logo–the "*GI*-profile". It is obvious how the *GI*-profile will look like for the above two limiting cases: for *GI*=2 bits/site each position accepts only one nucleotide ("A" for example) and the frequencies vector is (1, 0, 0, 0); and for *GI*=0 bits/site the vector is (0.25, 0.25, 0.25, 0.25), with no biases in frequencies. Hence, larger biases in frequencies signify higher information content.
We do not need to explicitly specify a genotype-phenotype mapping, but can investigate the influences of population properties, such as mutation rates, population size, fertility, recombination, and so on, on the *GI* and hence on a species' capacity for complexity.

Previously, we investigated a mutation-selection balance for varying genome lengths (Shadrin and Parkhomchuk 2014). Under the assumption of a slowly varying *GI* density and a sufficiently high (>1 per genome per generation) mutation rate, we explained Drake's rule and the "molecular clock" (Kumar 2005) phenomenon. We predicted the clock rate dependence on generation time (there are empirical observations of this dependence (Avise et al. 1992)) and explained how weakly and strongly conserved genes can have constant but different clock paces. Both of these phenomena can not be (as easily) explained by conventional theories (Schwartz and Maresca 2006). We showed that population size has no influence on *GI* storage in equilibrium conditions. Smaller size leads to more fluctuations around average values of biases, however these fluctuations cancel each other out, so that the total *GI* is not affected. Below, we assume constant population size and ignore the fluctuations, unless mentioned otherwise.

Maintaining larger biases puts a higher genetic load on a population. For example, any mutation in a site with *GI*=2 bits is lethal by definition. In contrast, no genetic deaths are needed to maintain a site with *GI*=0 bits, as it happens automatically. The lower the *GI* in a position, the less often a mutation will cause a genetic death, because the larger fraction of random mutations will increase the *GI* in a position. However random mutations still decrease the *GI* on average, hence the lower the mutation rate, the higher the *GI* can be. Reciprocally, a higher selection intensity—*i.e.* a larger amount of genetic deaths--increases the *GI*, which in turn increases the prevalence of beneficial alleles; so, for example, a higher fertility at a constant population size increases *GI*. However the means of *GI* increase bear physiological costs, be it the costs of DNA repair, or increase of fertility. Therefore, these trade-offs result in a balance, which produces the *GI*-profile.

There is no need to assume any particular genome size to grasp the basics of the proposed model. Therefore, we have found it useful for simple *gedanken* experiments, to use a genome with only a single site, dynamically representing it as a 4-vector of nucleotide frequencies in a large population, even though a single site is obviously not enough for recombination modeling. Selection is acting on the site according to the predefined selective weights of nucleotides, *e.g.* if "A" is the most preferable nucleotide, it would have the highest weight, and so on. With a sufficiently high mutation rate and a large population size, the site in a population will be occupied not only by "A"s, but also by other nucleotides according to their weights and available selection intensity (amount of genetic deaths). In a sense, this is a "weak" form of selection (which is unable to completely weed out detrimental alleles for a site), which we suggest is more ubiquitous than "strong" ("fixating") selection (Shadrin and Parkhomchuk 2014). A pure form of strong "fixating" selection will only be found for lethal sites, otherwise suboptimal nucleotides in a site can be observed with non-zero frequency in a general population. Then, the larger bias of nucleotide frequencies (higher *GI*) implies that the larger fraction of the population has the optimal nucleotide (*e.g.* "A"), so that this population is more fit, on average, than a population with smaller biases. This naturally holds true also for genomes with multiple sites, and it is useful to keep in mind that this higher *GI* population has a potential of being more fit in general, but not due to a specific locus.

Now lets consider a few sites in a genome, all with the same weights ("A" being the best allele), with few mutations per generation. In this case "perfect" genomes (*e.g.* "AAAAA") would be rare, in general, however the frequencies of "A"s in all sites will be increased. Then consider a "large" genome with the same few functional mutations per generation. In (mutation-selection) equilibrium, sequence statistical properties are homogeneous along the genome, so it does not matter where we drop these few mutations: selection can not "see" a genome size and can not distinguish where mutations were dropped, it "sees" only the cumulative effect on fitness of these few mutations, and acts correspondingly. So we can drop them, for example, in the first 100bp of the genome. However after mutation-selection round, statistical properties of these first 100bp will not change, since we are at the equilibrium, so the selection and fitness dynamics in equilibrium would be the same if we mutagenize only the first 100bp every time, and the rest of the genome being a passive hitchhiker. The genome size does not matter for selection dynamics, the dynamics is determined by the number of functional mutation per generation. After all, no one doubts that deleterious alleles are persistently lurking in a general population, *e.g.* Mendelian disorders, low penetrance and polygenic diseases and predispositions, and a population has some spread in fitness in general. It seems that the important novelty of our model is that we can determine this (potential) spread unambiguously from population parameters, along with the absolute value of "fitness" - total *GI*. Actual populations might not properly reveal this spread and potential variability, for example after a recent bottleneck or due to a small population size. However the mutational properties and fitness capacity of such population "slices", would be the same as for "large" (typical set) populations, so the recombination effects play the same role for them.

Up to this point, our considerations did not require the concepts of "information" or "complexity"; instead, one could deal only with "frequencies biases" and "fitness". However, there are some further points which justify the introduction of the information and complexity concepts: "fitness" traditionally is a relative function that is individual-specific. The "capacity for maintaining biases" is a population (species) property: a species' "channel capacity". It provides an absolute value which is interpretable as the total information content or complexity of a species and implies a potential and abstract "fitness" which is divisible (among loci). Such a "potential fitness" can be distributed differently (and transiently--no fixations necessary) among loci, depending on environmental demands, while traditional "fitness" is a single property of a genome (individual). The "capacity for complexity" can be also compared to "memory" (about an environment). It can be advantageous to have larger memory, however it is not "fitness" in conventional sense.

Changes in environment are equivalent to changes of the selective weight (and correspondingly *GI*-profile), and here investigate only constant weights, because it is natural to fully grasp simple scenarios first. If fluctuations of environment do not cause population size collapses/expansions, the demonstrated advantage of recombination is unaffected (total *GI* is unaffected), while parthenogenesis might be helpful otherwise, as we discuss below. Technically, recurring selective sweeps (*e.g.* due to environment changes) do not disturb the effects we describe. During the rise in frequency of some particularly beneficial allele, selection would react on accumulating "hitchhikers" in a similar way as in equilibrium (*i.e.* minus the fitness offset of this allele). In some models, it is admitted that such allele might bring also deleterious hitchhikers to fixation. In terms of our model such phenomenon can be reformulated: after the fixation, due the maintenance of new, increased frequency of this allele, (average) frequency biases in other sites will be inevitably decreased, *i.e.* the contribution to fitness from other sites decreases, due to the "channel capacity" limit. For brevity, we do not consider selective sweeps here, as if they all settled down.

For example Hill-Robertson effect, which is equivalent to Fisher-Muller model (Felsenstein 1974) critically depends on the presence of selective sweeps. When two (or more) unlinked beneficial alleles are rising in frequencies in a population, recombination might do some good by bringing them together. However it relies on a number of specific assumptions: clashes of highly beneficial mutations are frequent and background *de-novo* hitchhikers' effects are negligible in comparison with the fitnesses of driving alleles. However, should we conclude that in the absence of selective sweeps, in an equilibrium population, the recombination is useless? As we argued earlier, the number of sweeps is limited due to Haldane's arguments, so such *ad-hock* scenarios do not feel like an universal explanation for recombination.

Any species has a limited capacity for complexity, expressed here as genomic information, *GI*. Borrowing from information theory, we call this limit the channel capacity of a species. It is determined by mutation rate, genome size, fertility and other model-specific parameters, however not inasmuch by an environment, which is defined by selective weights distribution. If we change the weights, the *GI*-profile will be reshaped, but the total *GI* will not be significantly affected. For the above example of short genomes, with few functional mutations, we can play with the selective weights all we want, but it is easy to see, that if mutagenesis and selection intensity (*e.g.* fertility) are fixed, then we can not increase population average fitness (or, more accurately, the frequencies bias–prevalence of advantageous alleles as quantified by *GI*). By modifying weights we can not increase the frequency of perfect genomes "AAAAA"–they are just not present in sufficient amounts determined by mutation-selection balance (killing off all non-"AAAAA" would mean fast population extinction, as with few mutations per genome the progeny of "AAAAA" genome will unlikely contain "AAAAA"). Of course, we are able to introduce a skew: to increase the frequency of "AAANN", by increasing the weights of first three "A"s, but then frequencies of "NNNAA" and other advantageous combinations must decrease. A positive effect of increasing the *GI* and organismal complexity can be the increase of efficiency and fitness. However there is a peculiar and somewhat recursive interplay: there is a complexity maintenance burden (thus a decrease of efficiency), and there can be an increase of operational (survival) efficiency due to higher complexity. But eventually, complexity can increase if and only if the net efficiency increases. By analogy with information theory (IT), the increase of operational efficiency requires an increase of (algorithmic/hardware/organismal) complexity.

With all other phenotypic traits being equal, a species with increased energy efficiency will outcompete a less efficient species in an energy-limited niche, as they can achieve higher population density. Reciprocally a species with higher capacity for complexity may have an energy efficiency equal to that of a less complex species, but can devote extra capacity for complexity to enhance the efficiency of some other phenotypic features, gaining again an advantage (though it should be possible to reduce any phenotypic efficiency to energy efficiency, see mimicry example below). However, in contrast with many other complex phenotypic features where the operational efficiency gain is subjectively transparent (like elaborate mimicry), the increase of efficiency due to the maintenance of recombination is far from being apparent. To the contrary, in general it is admitted paradoxical in a number of aspects. Notably, the increased efficiency of mimicry (for example) can be also objectively expressed (formally quantified) in terms of channel capacity and thus energy efficiency: Assuming that the improved mimicry allows for longer average reproductive life span, it allows more progeny per individual, so selection on progeny becomes more intensive, which results in the increased channel capacity for the population. It can be said that mimicry, or any other phenotypic feature (which requires maintenance) compensates its maintenance burden by improving efficiency and channel capacity, justifying existence of itself. We can currently observe only those genetic systems which persisted and managed to ascend through billions of generations. From this perspective, the central "objective" of a genetic system is its survival--persistence in time, or transmission of its genetic information through generations without degradation (at least). Applying this rule to recombination we just need to show a mechanism provided by it, which increases capacity for organismal complexity by improving the fidelity and bandwidth of the *GI* transmission, allowing to maintain larger frequencies biases. This logical inevitability of such role for recombination solves the problem on conceptual level. Below we illustrate it with more specific minimalistic model implementation. Naturally other, more complex implementations are possible.

**Simulation**
There can be an infinite number of non-equlibrium but transient evolutionary scenarios as opposed to the inevitable equilibrium stage, hence for the start we consider a simple stationary, equilibrium population (Fig. 1) similar to the one in (Shadrin and Parkhomchuk 2014). We assume that a limited niche is populated densely, *i.e.* the population can not grow because it reached a balance with available resources. Hence when an individual dies, it is replaced with a newborn individual.

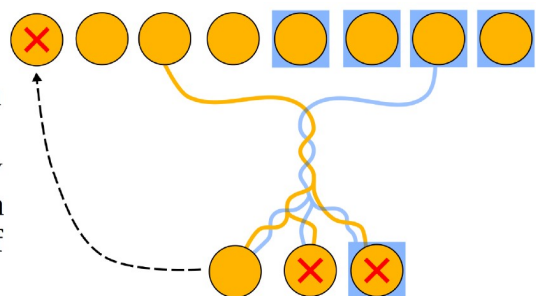

**Fig. 1.** Diagram of an iteration step. An individual removed from population is replaced by new individual selected from a litter. In sexual population half of individuals are males (blue).

To single out the effect of recombination, we tried to minimize the number of model parameters,

presented in Table 1, to reflect only the most general and critical steps of the replication-(recombination)-mutation-selection process in equilibrium. Fig. 1 represents a single iteration step of the simulation. If we consider only yellow colored individuals (and lines), we look at asexual reproduction, and when half of the population are males (blue) it is a sexually recombining population. The selection (fitness function) of progeny is based on a simple sum over a genome of four (AGCT) nucleotide weights for all positions (Eq. 3 in (Shadrin and Parkhomchuk 2014)), which is proportional to typicality. The specific formula (*e.g.* additive or multiplicative) for fitness does not matter, it should just provide an opportunity for selection to increase frequency biases, compensating the degrading role of random mutagenesis. For simplicity, the weights are the same for all positions (Table 1), because the actual differences in composition do not matter for the quality of the described effects. The selection of the initial reproducing individual was excluded, as it gives equal offset for all populations and is not important for demonstration of recombination effects (technically, it can also be reduced to selection on progeny). We consider three types of populations: asexual populations, sexual populations that recombine randomly and sexual populations that perform sexual selection. A random individual from the female population fraction is selected as a first parent and in an asexual population this is the only parent. The second parent for a recombining population is also selected randomly, from the male fraction of the population. And for a population with sexual selection, this random selection only considers those males with the highest weight (fitness). This increases the selection intensity for this population and quite predictably improves the fitness of the offspring and the channel capacity of the species. In the more interesting comparison of randomly recombining vs. asexual populations, they have exactly the same selection intensity, defined by the number of children produced, out of which one is selected to replace the removed individual. We found that it does not matter whether the removal of the individual is done by age, randomly, or by the lowest weight (fitness). The gender of the surviving child is defined by the gender of the removed individual (so the sex ratio remains constant in the course of the simulation). In a recombining population, there are only half the number of females. However, the effect of the reduced reproductive capacity--at constant population size--may come to play only when females are unable to compensate for natural deaths in a population, which seems to be a rather special (and non-equilibrium) situation. However, for most species a female produces significantly more than two children, so the reproductive capacity is not an issue in a constant population. As could be expected, the central role in the phenomenon is played by selection, acting on the results of recombination (selection on progeny).

The "two-fold" costs of recombination are apparent only when there is an expansion into an empty niche, without competition and limits on reproduction. With other things being equal, an asexual species would multiply at least two times faster than a sexual species with the same population size, but half of it being males and only the other half being females. Parthenogenesis does have an advantage in the rate of population growth, in those cases when it is allowed. If recombination was advantageous in all cases, we would not observe parthenogenesis at all. However, such an exponential growth will always be short-lived: for instance, if a population starting from a single organism was allowed to double merely 200 times, the resulting population size would be much larger than the number of atoms on Earth (~$10^{50}$). The expansion is followed by saturation at the carrying capacity. The population growth halts and the competition for resources starts. Thus, coming back to our model, the reproduction (renewal) rate is limited by resources in both populations and equals to the death rate. Hence, the niche becomes more complex, and the two-fold reproductive advantage disappears, while demands for efficiency and complexity increase. Now a (potential) reproduction rate is not an important parameter, as a niche can support only a limited number of individuals, while population density and hence efficiency becomes important.

**Table 1.** Parameters and pseudo code of single iteration step

| General population | Value |
| --- | --- |
| population size | 100 |
| genome length | 100 |
| nucleotide weights for fitness function | (0.65, 0.25, 0.1, 0.0) [AGCT] |
| per base probability of mutation | 0.05 |
| transition/transversion ratio | 2 |
| number of descendants in litter | 4 |
| **Sexual population** | |
| female/male ratio | 1 |
| probability of recombination | 1 |
| per base probability of crossover event | 0.1 |
| strength of sexual selection | 0 (no selection), 0.5 (2$^{nd}$ parent from top 50%) |
| **Pseudo code of single simulation step** | |
| rand = random float from 0 to 1<br>REMOVE random (or the least fit, or the oldest) individual from a population<br>SELECT a (pair of) parent(s) as described in text<br>FOR each child in a litter:<br>    FOR each consecutive genome position of a child:<br>        IF rand < per base probability of crossover event:<br>            switch to incorporating sequence from other parent (including current position)<br>        IF rand < per base probability of mutation:<br>            perform random substitution with given transition/transversion<br>SELECT the child with the highest weight in the litter and put him into the population | |

**Results**

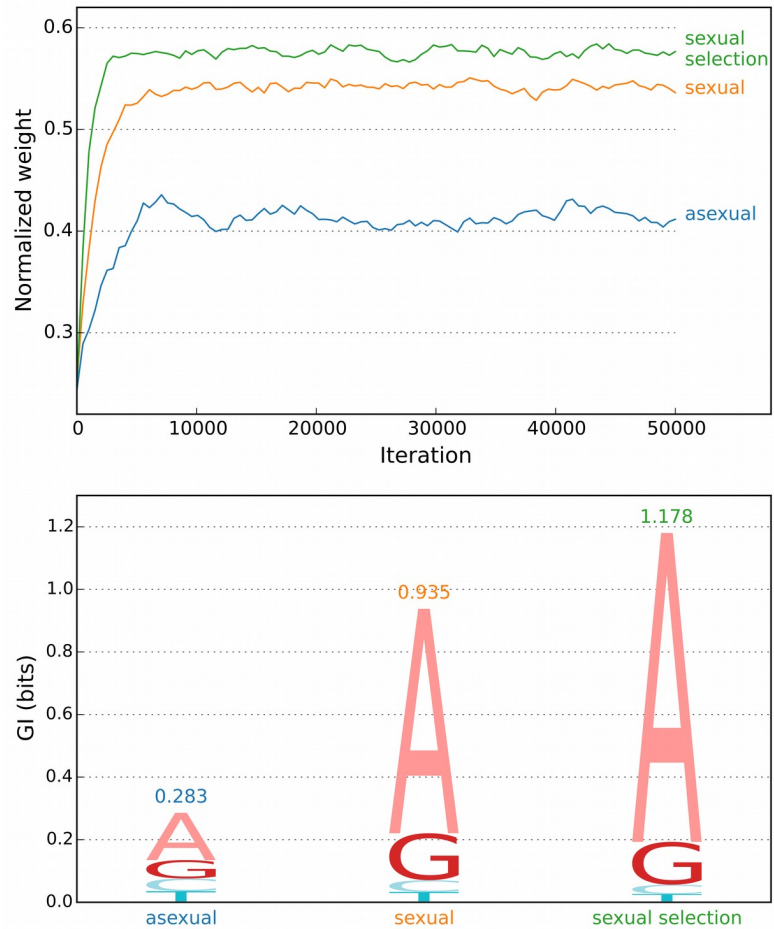

**Fig. 2.** Three types of populations evolving to their equilibrium state from random genomes. At the bottom, average biases of frequencies are shown over the last 40,000 iterations.

We wrote a simulation code in python (https://github.com/interCM/evolution) according to the pseudo code and parameters in Table 1. We performed a large number of experiments, with various parameters, and the results were robust and fast-converging, with no appreciable variability between experiments. Fig. 2 shows three population types evolving *in silico* to equilibrium states. The asexual population has the lowest average weight (fitness) and thus the lowest channel capacity and potential complexity. It can be compensated in the asexual population by lowering the mutation rate about 3 times(!) (Fig. 3, bottom, dashed lines). The equilibrium nucleotide frequencies averaged over all positions are represented in logo format at the bottom of Fig. 2. The weight is normalized to genome size, so it is the weight per position.

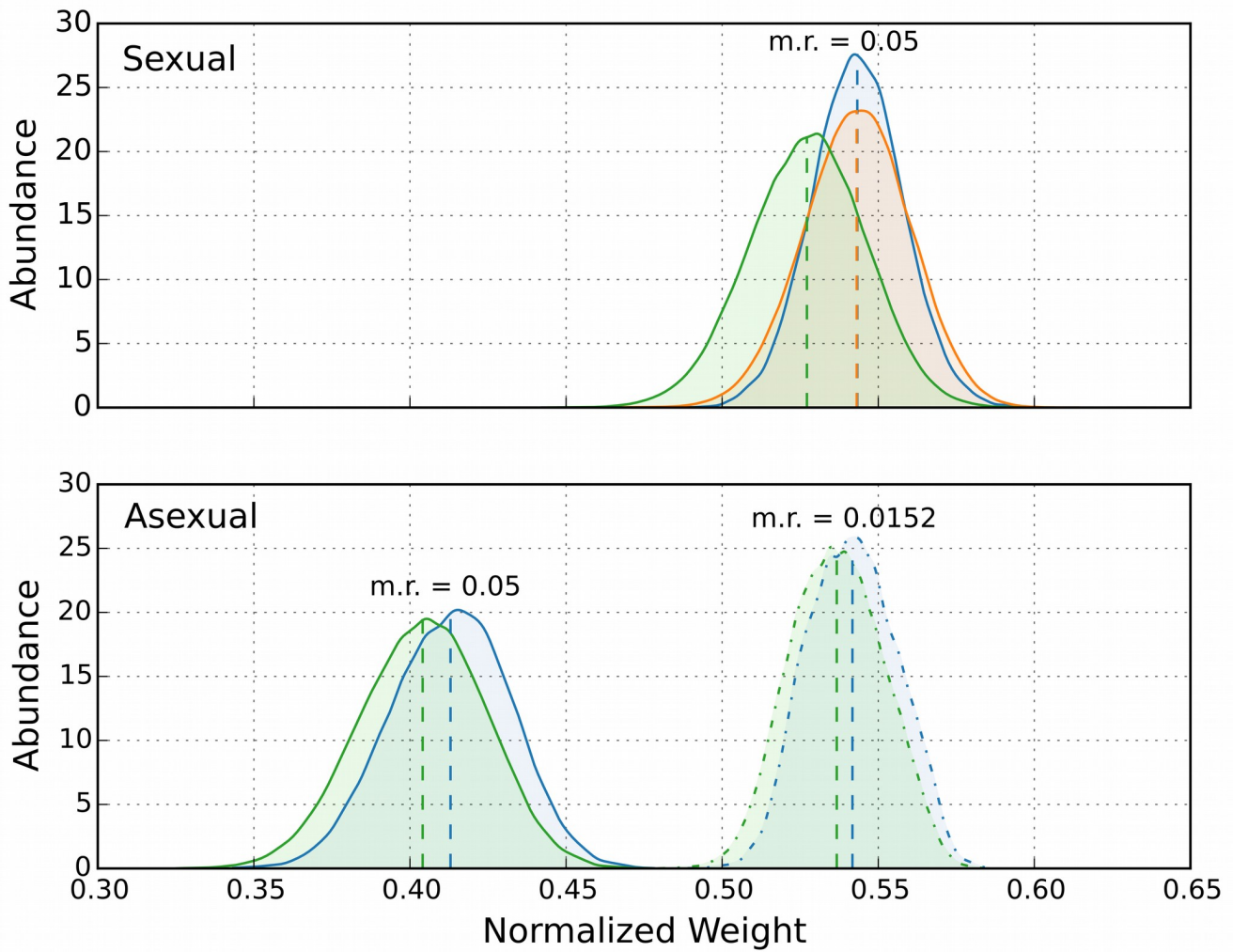

**Fig. 3.** Averaged distributions of weight (fitness) in progeny, after corresponding steps: blue – after selection, orange – after recombination, green – after mutagenesis (before selection). Top – sexual population, bottom – asexual, with the same parameters. Solid lines represent mutation rate of 0.05. The dashed distributions for the asexual population (bottom) result from the three times lower mutation rate (to ~0.015) in the same population.

For the above (Fig. 2) simulation we collected data for fitness distribution in progeny, accumulating separately the fitness values of progeny after each step of recombination-mutation-selection process. In equilibrium a population has some established fitness distribution, which is modified in a closed loop by mutation, recombination and selection steps. Fig. 3 separately demonstrates the averaged effects on the distribution of weight (fitness) in progeny caused by mutations, recombination and selection: During the recombination step, the average weight does not change, but its variance increases (orange curve is "wider" than the pre-recombination distribution–blue curve, Fig. 3), mutations decrease the average weight, shifting the distribution to the left (green line, Fig. 3), and then selection amplifies the right side of the distribution and suppresses the left tail, shifting it back to the right (blue line). Random mutations also increase variance in fitness, it is not that visible for the asexual population, however the green curve for sexual population is noticeably wider than the orange one (these effects are not very relevant for the demonstration). Naturally some of mutations are positive. The averaged effect of recombination is mathematically analogous to convolution, a "smearing" of the parents' fitness. So the recombination acts like mutations, but without shifting the average fitness, it is "mutagenesis" with 50% of positive mutations. Consider two recombining parents with equal weight (for simplicity), but with different sequence compositions. If we assume that there are no novel mutations, the average weight over all their (non-monozygotic) children will be the same, but due to the reshuffling of their sequences, they acquire variance in fitness: their fitnesses will be distributed below and above their parents' fitness. Thus, recombination alone does not change the average weight of progeny. Only at the selection step will the left tail of the progeny distribution be eliminated (or suppressed) and the right tail be emphasized. As a result of this selection on progeny, their eventual fitness can be higher than that of both parents. The same increase in fitness variance happens in any general population, so it is immediately apparent how recombination plus selection increases the average weight, in comparison with asexual reproduction. Recombination provides the extra right tail and at the selection step this tail is promoted, while the detrimental left tail is discarded. An asexual population with the same mutation rate (solid lines) is shown at the bottom of Fig. 3. It has a significantly lower average fitness and thus *GI*, because it lacks the extended tails provided to the sexual population by recombination. In order for the asexual population to have a *GI* (average fitness) similar to that of the recombining population, the mutation rate in the asexual population must be lowered to about a third (!), from 0.05 to 0.015 (Fig. 3 bottom, right dashed distributions).

It seems that the larger allelic differences in the case of two (equally fit, for simplicity) parents might produce larger variance in progeny fitness; thus we may speculate that the extended right tail of the progeny fitness distribution might describe the "hybrid vigour" phenomenon, while the left tail might partially hide from observations due to a drop in fitness that is so significant that it causes an early termination of development.

One of the main determinants of channel capacity is the mutation rate–with a lower mutation rate allowing for a higher channel capacity (Shadrin and Parkhomchuk 2014). For example, we doubt that the rate decrease towards human lineage is merely a coincidence: *e.g.* Fig. 1 in (Scally and Durbin 2012). However, a rate decrease also requires an increasing complexity of DNA repair or proofreading mechanisms, and thus a complexity burden. Also, too low rate would decrease evolvability. Therefore, in general, the rate is as high as possible for a given species and a given niche complexity (though there could be notable exceptions discussed below). Niche complexity is an important factor for organismal complexity evolution and maintenance: a tooth would not evolve by itself when there is no hard food available to bite; reciprocally, eyes would degrade in total darkness after passing some generations. If we describe a phenotype in a broad sense as an interface between a being and its environment (as perceived by the being), then it is clear that the interface, as a border, portrays both of them. In an optimal situation, the complexity of a niche is matched by complexity in a corresponding species, or "life is what it is because the environment is what it is" (Pike and Scott 1915).

Fig. 3 shows that the effect of decreasing the mutation rate (increasing DNA repair efficiency) on the channel capacity is comparable to the effect of introducing recombination; that link might help us to explain some confirmed long-term asexuals later. This connection between the roles of DNA repair and recombination represents a peculiar functional "convergence" (of inter-individual and intra-cellular mechanisms) since the recombination mechanisms on the molecular level were likely branched out of molecular machinery for replication and repair of DNA, *e.g.* a repair by sister chromatid exchange and homologous recombination. Hence, there is no logical gap in the "motivation" of the evolutionary ascent of recombination: the basic repair-like role of recombination was simply honed and extended further for inter-individual DNA exchange. A notable difference to intra-cellular repair is that the repair-like effects of recombination are only revealed through the coupling of recombination with selection acting on its results (progeny). And recombination acquired other useful properties in comparison with intra-cellular repair: increasing the genomic complexity through decreasing the mutation rate alone might have a drawback due to decreased evolvability; in contrast, the "outsourced repair" via recombination seems to address two issues: it increases variability in a population (higher evolvability) and allows for significantly increased mutation rate (a lower repair burden and an additional increase of variability) with the same average fitness (Fig. 3).

**Discussion**
The increase of biases in the equilibrium frequencies of alleles due to recombination is a simple and robust effect of recombination. Since we suggested a mechanistic model to quantify fitness, genomic complexity and organismal complexity by these biases, we naturally presume that the effect can be interpreted as an increase in channel capacity or capacity to maintain complexity.

The proposed model is practically parameter-free: the mutation rate is not an independent parameter because of Drake's rule (Shadrin and Parkhomchuk 2014), while *GI* density (<2 bits/site) and genome length have no qualitative influence on the described phenomenon. Linkage effects obviously play no role in the equilibrium state, so we can freely scale genome size, mutation rate and population size to their realistic values, and will observe the same phenomenon. We experimented also with larger genome lengths, and found the results invariant, so we chose 100bp genomes for computational convenience. In contradistinction to many previous models, there is no qualitative dependence of the effect on population size (which affects only the observed magnitude of stochastic fluctuations around the equilibrium trend) or the mode of reproduction (overlapping/non overlapping generations). We don't need to postulate any epistasis or synergy, though a reasonable inclusion of such effects will not alter the basic mechanism. The only critical features for this simple model to be applicable are: 1) a sufficiently high mutation rate (>1 per genome per generation) to maintain a persisting variability in a population (which seems to be ubiquitous empirically, and obviously recombination is useless in a monoclonal population); 2) a densely populated niche (which is inevitable due to exponential growth periods), resulting in a constant population size and reproduction (and death) rate; and 3) selection acting on overproduced progeny, selecting the most typical ones, as calculated by the simple sum over nucleotide weights.

We suggest a model that has an important advantage in comparisons to previous models, which correctly admit the burden of maintaining increased complexity in sexual populations but lack consistent units of measurement, and the channel capacity notion. We propose, that complexity can be formally expressed in bits: the amount of genomic information passing from generation to generation for a given species. In contrast, traditional models try to compensate for the increased complexity burden by quantifying the "goodness" or "badness" of particular mutations. To this end, they assume a certain fitness associated with a certain allele at a locus and for interactions of different alleles and loci, they create fitness functions to describe them. However, both the increased burden to maintain extra complexity and the capacity for efficiency improvements it may create, should be consistently represented in bits. In doing so, our framework (Shadrin and Parkhomchuk 2014) can naturally account for fitness on a genomic level, instead of having to separately consider each allele, each locus and each of their interactions.

If we ignore the changes in species complexity (its genomic information), then we can not even tell if some molecular evolutionary process, which we model, is "progressive" or "regressive" evolution; if we are unaware of complexity dynamics, on the unobserved genomic complexity scale such process might look like an opportunistic random walk. This does not seem very biological, making the direction of "evolution" rather vague. Usually, evolution is defined as an increase of fitness. However, without the formalism for species complexity evolution, we cannot tell if the increase of some (arbitrarily chosen) fitness, in a given model, results in progressive or regressive evolution: does species complexity increase or decrease? That is clearly a central question for the long term evolutionary destiny of a species. Many authors tried to address the "long term advantage" of recombination, but were lacking the proper formalism to trace complexity evolution, which in our opinion, is the long sought "long term advantage".

Another related critical point of many evolutionary theories is that they consider mutation fixation as an elementary step of evolution, *e.g.* (Desai and Fisher 2007; Hartfield and Keightley 2012). This introduces strong dependencies on the absolute population size. Correspondingly, these models studied the influence of recombination on fixation process. To the contrary, in our model fixations *per se* play no role in complexity evolution: they are a transient phenomenon, which can be averaged out. As noted by Fisher, if a variant is not lethal, it will be present at some frequency in a sufficiently large population, or equivalently, it will appear *de novo* at some frequency in a smaller population. Thus, a proper descriptor of a non-lethal variant would be its average frequency over time and the absolute population size can then be excluded from evolutionary dynamics. Lethal sites constitute a small fraction of a genome (otherwise the molecular clock would not function), and due to their lethality, and hence invariability, their contribution to evolutionary dynamics is questionable. It seems that (conceptual) fixation on fixations in traditional theories was provoked by a concept of a "reference genome" (or "wild type") for a species. A "reference genome" alone gives no clues of species complexity or total genomic information (G- and C-value paradoxes), because it lacks the information about degree of sequence conservation (hence its informational value), which can vary widely. Hence, we suggested that a better representation of a species as a whole (and its fit to its environment) might be a cloud of genotypes--a "typical set" that contains all possible genomes for a given species. Then, any "reference genome" is merely an equal member among many others (Shadrin and Parkhomchuk 2014), and species evolution is a modification of the typical set. Our simplified but general formalism shows, that *GI*-profile can describe the conservation profile and the size of corresponding typical set can be used to estimate species complexity and genomic information, in a meaningful way.

It can be loosely said that all the initial hypotheses are involved: recombination indeed helps to remove negative mutations and to promote positive mutations, and this is happening because of the increase in variability (Fig. 3). The removal or promotion of mutations are two sides of the same process--fixation, which in traditional models necessitates the introduction of the absolute population size. However, the crucial nuance in our case is that allele frequencies would be constant in a "large" equilibrium population (which is synonym with a typical set). Nothing would be removed completely or promoted to fixation. However, in a smaller population, fixations would happen due to drifting around corresponding average frequencies, but would still have no influence on the total *GI*. So the function of recombination is not helping to "remove" or "promote" in the classical sense, but rather to increase biases in (average) alleles frequencies (Fig. 2, 3), thus increasing *GI* and channel capacity, and there is no need to deal with the absolute population size. In this case, the biases (of average frequencies) of good and bad alleles increase in corresponding opposite directions and fixations *per se*, happening due to the limited population size, play no role in complexity evolution. Roughly speaking, the fixation by drift of a bad allele is on average compensated by the fixation of a good allele, and changes in population size affect the frequencies of both fixation types equally, without affecting the total *GI*. Naturally the change of biases (or population size) will lead to changes in fixations (drift) dynamics, however this is irrelevant for our approach here. There is no sense to debate about which of the mechanisms of recombination is more "important": the removal of bad variants (mutations purging), or the promotion of good variants, as it is the change of biases which leads to the change in channel capacity. And by definition, a bias is a shifted balance, in our case a balance between good and bad alleles. A recombination advantage can be also loosely formulated in terms of fixation or a reference genome. It can be said that in a small recombining population, where a reference genome can be "defined", more advantageous mutations are fixated at any given moment, in comparison with asexual population. That's why with other things being equal the recombining population can have higher *GI* and fitness. However, in the course of "neutral" drift, these advantageous mutations will be replaced stochastically by other advantageous mutations, so they are not really "fixated", though their prevalence over fixed negative mutations in recombining population will be consistent. Except for the lethal sites, any fixation is a temporary, fluctuating, transient state of affairs. And to reveal the prevalence of positive variants, it is necessary to have a species-level (not individual level) genomic outlook available, which we provide with *GI*-profile and typical set.

In a sense, we presume drift to be an "ergodic" property within the typical set as a "phase space". This is equivalent to the assuming that genomes in a limited stable population under the process of "neutral" (*sic*) evolution, will eventually visit all possible genomes of a given species (its typical set). A "large" population then represents a complete typical set ("phase space") and a sufficiently long time average over a "small" population is equivalent to the "large" population (substituting dependence on the population size with dependence on time) . This opens an opportunity for applications of the developed mathematical apparatus of ergodic theory in population genetics. We suggest that instead of studying how drift dynamics depend on population size and other parameters, one can study the properties of the typical set as a container for complexity, a species property that can be optimized. In contrast, the population size cannot be optimized significantly, as it is not defined by a species *per se*, but rather--in the case of the saturated niche--by available local environmental resources. These can be patchy and opportunistic, and the only role for a population is to fill the niche efficiently. For example a large homogeneous niche would allow for the larger effective population size, however there is no empirical data, which indicates that such niches are conductive for evolution.
Naturally a "topology" ("connectivity") of a typical set can be complex in general, in some sense it is analogous to the "fitness landscape" notion.

Another aspect worth considering is that for two species with the same total *GI* the one with lower *GI* density (and thus a bigger genome) can presumably benefit from recombination more: With more entropy at individual genome positions, its typical set will be larger. This implies a larger population variability, which is a substrate for recombination action. Hence, recombination is more beneficial for genomes (genomic regions) with lower *GI*. Assuming the latter arguments are true, we can expect that more complex species, tending to resort to low *GI* density evolutionary strategies (in contrast with simple species, which usually have conserved genomes with high *GI* density), will try to intensify exploitation of recombination advantages. It could be that lowering of *GI* density and enhancement of recombination intensity have occurred in parallel in the course of evolution.

Since sexual selection can significantly increase *GI* (Fig. 2), beyond the necessities of bare survival in a niche, it is natural to predict that it will allow the emergence of complex phenotypic features devoted solely to sexual selection, for example brightly colored plumage of many male birds, fishes and many other behavioral and social phenotypic features. In higher animals, females bear significant reproductive complexity load, hence, predictably, males are better suited for being more "colorful". As noted by Darwin, such features seem to provide no advantage for routine survival, but arose solely due to sexual selection. However, we have to stress that this is an incomplete "explanation"--to make it complete, we have to show *how* sexual selection provides additional capacity to maintain extra "unnecessary" complexity, without hurting the complexity maintenance of "bare necessities". Otherwise, one has to explain why asexual species--without the burdens of recombination and various unnecessary features associated with sexual selection--usually can not outperform the sexual ones. Hence, sexual selection allows building up a kind of "cultural" layer of complexity that is unnecessary for bare survival.

The demonstrated advantage of recombination is inherently multi-generational. The switch to parthenogenesis is equivalent to significantly increasing the mutation rate (Fig. 3). Thus, the species complexity and operational efficiency will start to drop gradually. At the same time, they acquire an immediate "two-fold" reproductive advantage, although we repeat that this advantage could only play out in "unsaturated" niches. This explains the difficulties with its experimental study and theoretical modeling (without regard to complexity evolution), and evolutionary short lifetime of lineages converted to pure parthenogenesis. After the conversion the complexity and thus fitness will keep degrading through generations due to the increased mutations load, unless these species manage to drop mutation rate significantly. Many simpler species can readily loose the ability to recombine (Tucker et al. 2013); if these lineages were competitive we would observe large number of them, however pure long-term asexuality is scarce. On the other hand, facultative parthenogenesis can be obviously useful in cases of population expansions, or in niches with lowered complexity, *e.g.* "disturbed habitats" or extreme, "marginal" conditions with decreased biodiversity (Vrijenhoek and Parker 2009).

It is easy to see, that the common denominator of an "empty" niche or "marginal or disturbed" habitats, where asexual reproduction prevails more frequently, is the lowered niche complexity, *i.e.* less competitors, parasites and so on. In these conditions with lower demands for efficiency, asexual reproduction, without additional burden of recombination and having a reproductive advantage, might be beneficial for capturing such niches. For marginal habitats with very limited resource density (resulting in a low population density), the costs of finding a mating partner can be too high, so parthenogenesis might also be competitive in such equilibrium cases. However, the survival on an evolutionary peak in a rich competitive niche, requires the maximization of species complexity and evolvability, which is attainable via recombination. Then we can expect that on the scale from simple to complex species, recombination becomes more necessary the more complex a species is. Probably, the burden of retaining a viable option for parthenogenesis was not worth it for highly evolved species. As highly evolved species acquired more independence from environmental fluctuations, they became less prone to population collapses/expansions caused by these fluctuations, and the utility of parthenogenesis diminished. For highly evolved species, facultative parthenogenesis could be selected against in the long run: indulging in parthenogenesis bears the risk of loosing the ability to recombine or degradation of its fine-tuned properties ("use it or lose it"), which implies extinction; the higher the species complexity, the higher the risk and the faster the extinction. There is a direct experimental example that sexual reproduction prevails in heterogeneous environments in comparison to homogeneous ones (Becks and Agrawal 2010). However in that case the effect was probably not due to the species complexity decrease and "extinction" because of the increased mutation load (too few generations had passed), but likely a "second-order" adaptation ("anticipation") of species to switch between sexual and asexual modes of reproduction in corresponding environments. In this study the rate of sex was decreasing in all laboratory conditions for field-collected (complex niche) samples and the authors did not find any evidence that genetic drift (and thus population size) plays any role in recombination advantage. For the same reasons of decreased niche complexity in agricultural environment we would expect parthenogenic reproduction success therein (Hoffmann et al. 2008).

Another phenomenon naturally explained by our model is, that many bacteria and other organisms capable of parthenogenesis actively resort to recombination in nutrient-limited conditions. A development of such kind of "foresight" or "anticipation" should not surprise us (Rosen 2012; Nadin 2014), as the predictive capability (*e.g.* predictions based on previous data) is the central theme in algorithmic complexity (in Kolmogorov's sense) or channel capacity optimization by data compression in IT. Evolution towards independence from external conditions (Pike and Scott 1915) implies some "learning" and "understanding" of these condition, with increasing anticipation of external influences and counteracting them with proper means. Eventually, this manifested in learning capacities and planning (projecting, forecasting and so on) preoccupations of a rational mind.

From these general considerations it is easy to predict that higher biodiversity (higher niche complexity) would lead to higher energy harvesting efficiency (Schneider and Kay 1994; Cardinale et al. 2012). Ecological balance models mainly focus on the balance of resources, such as predator-prey models. Complexity evolution considerations might add an interesting "anti-monopoly" dimension to ecological (and socio-economical) balance models: for example, a locally successful species, which was able to eliminate many competitors, will "suffer" from the decreased niche complexity and subsequent degradation, eventually becoming susceptible to extinction and vulnerable to "less successful" neighbors. In a sense, if wiping out competitors (or prey) might be suicidal in the long term, the anticipation of this phenomenon could count as "altruism" or "humanism".

*Bdelloid rotifers* seem to have stuck to parthenogenesis for millions of years and are perceived as an "evolutionary scandal" (Welch 2000; Welch et al. 2009). However we showed that the effect of introducing recombination is analogous to increasing repair efficiency and lowering mutation rate. Hence, we can promptly hypothesize that there is something unusual about their repair efficiency. Indeed, these species have the unique ability to survive dehydration at any point of their life cycle. Dehydration induces a lot of DNA damage, *e.g.* double strand breaks (Hespeels et al. 2014), hence the repair machinery of *Bdelloids* must be ready to repair a lot of damage at any time. Therefore, we can assume that this machinery also actively repairs damage under normal conditions, so the errors are being repaired beyond the usual necessity to maintain the complexity required by their niche. It was also shown that *Bdelloids* are extremely tolerant to radiation, likely because of their repair efficiency (Gladyshev and Meselson 2008). Hence, we can hypothesize that their mutation rate and repair activity are determined not by their niche complexity, as in most species, but they are honed by the demands to survive those mutational shocks during the dehydration. Hypothetically, they have plenty of potential channel capacity but their simple niche does not require further increase in complexity. In this situation the repair activity would passively degrade in ordinary species, down to the point where the capacity matches the niche complexity and recombination would become potentially useful again; however this is not happening in *Bdelloids* for the above reasons. There is no point to maintain sexual reproduction for them, as they gain nothing useful from it besides the burden. Also the population recovery after dehydration state seems to represent an expansion into a simple niche. It would be interesting to evaluate their performance in more stable and rich environments, the prediction would be that they will lose competition to close species with sexual reproduction (*e.g.* other rotifer species) and their mutation rate would passively increase without the selection on advanced repair; but to a certain point this rate increase will not affect their phenotype and performance.

Another example of supposedly ancient asexual species are ostracods *Darwinula stevensoni*, though some males specimens of *Darwinulidae* were found (Smith et al. 2006). This species were reported to have decreased mutation rate (Schon et al. 1998). However the reasons for this are less obvious in comparison with *Bdelloids* and are still "intriguing" (Van Doninck et al. 2004). Notably, this species supposedly have increased environmental tolerance to salinity, temperature and hypoxia in comparison with close relatives. Ostracods eggs can also survive dehydration. It might be that some of this factors impose selective pressure on advanced DNA repair mechanisms similarly to *Bdelloids*, alleviating the need for recombination and making ostracods prone to form relatively stable asexual lineages.

Another interesting experimental system with a potential for multi-generational experiments are thrips, as they can readily form both sexual and asexual populations (Kobayashi et al. 2013) (*e.g.* one could try to compare long-term mutation accumulation rates). Yet another promising candidate for empirical evaluation of the proposed model are sex chromosomes, in particular the mammals' Y chromosome, which is not recombining. As stated in (Graves 2006) "One of the biggest mysteries in biology is what drives degeneration of nonrecombining regions of the Y, and why positive selection for male-advantage genes does not stop it.". From the point of view of complexity maintenance its degradation is rather predictable, as it has the lowered channel capacity. Y chromosome is known to have few times higher mutation accumulation rates in comparison with other chromosomes (in full accord with our "predictions"), and genes residing there tend to degrade. However, more sophisticated analyses are required to quantitatively fit our hypothesis there, because the Y chromosome is restricted to the male germ line, where more cell divisions might occur per meiosis, and it also may lack some aspects of homologous recombination repair present for paired chromosomes. It is inefficient to use it for the storage of "normal" genes, so it is more of a marker for sex determination purposes.

On a philosophical side, considerations about motivations for increasing complexity (the drive to optimality or efficiency) might help to bridge the gap between reductionism and more "metaphysical" lines of thoughts such as emergentism, spontaneous self-organization, self-organized criticality and so on. Previously we introduced a limit on the amount of genomic information, and here we showed the importance of another related limit: the limited resources in a saturated niche (hence a permanently "critical" condition there). In mathematics we get used to routinely operate with infinities and open-endedness, *e.g.* "all" natural numbers and so on. The question is whether we can build models of complexity evolution without respect to (physical) constraints, or maybe the constraints are the very reason for complexity unfolding in physical (and mental as part of it) world, as opposed to Platonic world of all possible forms and algorithms (which is as "real" as "all" natural numbers), where the complexity seems to be infinite but static. An "efficiency" has a meaning only with respect to some constraint (*i.e.* resource-efficiency). One can speculate whether the axiomatically postulated and ubiquitous in physics Maupertuis' principle of "thriftiness" is intricately related to the "naturally occurring" thriftiness role in the evolution of biological systems. Simple objects (*e.g.* isolated elementary particles) implement simple algorithms for efficiency, *e.g.* "least action", while complex compound objects implement more complex algorithms of collective behavior. For example in a classical emergent phenomenon such as Benard cells, the ordered structures emerge supposedly to act as more effective energy dissipaters, and hypothetically organisms and ecosystems have analogous "motivation" (Schneider and Kay 1994). What is then a "creative" (emergent and so on) event or an increase of algorithmic complexity in a system with an energy gradient ("dissipative")? We might speculate that a system somehow pulls out (of Platonic world) an algorithm for increasing efficiency and physically implements it. This act of "pulling out" looks mysterious, however it happens reproducibly though with some degree of chaos (*e.g.* Benard cells and many other examples) in properly prepared conditions. This implies that the link between the algorithm and appropriate conditions was pre-existing, but unobservable before the conditions were met, which looks like an interesting variant of preformism, and the link resembles a "*categorial novum*" (Poli 2011). The algorithm "existed" in some (Platonic) sense, but the *only* way to implement it, or "become aware of", "discover" it, is to reproduce the proper physical (mental) conditions; *in silico* and *gedanken* experiments count as variants of physical implementation. Hence the alleged unpredictability and irreducibility of emergentism are justified in "real life": with limited physical resources it is impossible to enumerate all possible physical conditions and thus deduce and learn all "natural laws" (implementable algorithms). The numerous puzzling and seemingly improbable cases of convergent evolution (*e.g.* flight in birds, mammals, insects and so on) might be a manifestation of the reproducibility of emergent phenomena. If an emergent phenomenon happens to play a purposeful role as an efficiency enhancement (such as Benard cells), then genetic variation honing it will be selected for, which is reminiscent of "evolution without natural selection" (Lima-de-Faria 1988; Kull 2014). The accent (in causation) here is that the "creative" or "adaptive" feature, increasing complexity by breaking symmetry, first appears as an act of emergence in a sufficiently complex system, but not due to a specific mutation *per se* (hence the feature can re-appear independently in different genetic systems); then natural variability due to mutagenesis picks up, amplifies and fine-tunes the feature (so selection is still there but it merely plays a secondary, maintenance role). Such scenario suggests that progressive evolution is a mixture of gradual improvements by selection and discontinuous events of capturing novel emergent phenomena, which appear when a system (including environment) reaches certain complexity threshold. It might be that mind's "creativity" is a skill of merely preparing the right (often quite intricate) mental conditions (trying to "solve a problem", constituting a "conscious work" in terms of Poincaré), and then a link to an efficient solution might emerge instantly ("unconscious work") (Poincaré 1982), implementing the solution/algorithm physically/mentally. The reproducibility of a mental creative act provides an opportunity for learning, though does not guarantee it, as some previously acquired knowledge base, efforts and skills for preparing the right conditions are required. Then there are the clear differences (well-known to educators) in learning by "memorizing" and learning by "true understanding" (recreating proper conditions and experiencing an emergent act, in our terms), with possible consequences for AI research.

Hopefully the improvement in understanding of genomic complexity evolution will enhance our capacity for preventing and combating evolution of pathogenic species, pests, and common polygenic diseases and predispositions; as well as manipulating artificial selection, cloning and genetic engineering, which already play a tremendous role for humankind, simplifying its life *en mass*.

**Acknowledgments**
The authors are very grateful to David Laehnemann for numerous comments and suggestions. D.P. gratefully acknowledges funding by German Center of Infection Research (DZIF)